\documentclass[aps,twocolumn,groupedaddress]{revtex4}
\usepackage{graphicx}
\usepackage{natbib}
\usepackage{dcolumn}

\begin{document}

\title{Natural Orbitals and BEC in traps, a diffusion Monte Carlo analysis}
\author{J. L DuBois}
\author{H. R. Glyde}
\affiliation{Department of Physics and Astronomy, University of
Delaware, Newark, Delaware 19716,USA}

\date{ \today}

\begin{abstract}
We investigate the properties of hard core Bosons in harmonic
traps over a wide range of densities.  Bose-Einstein condensation
is formulated using the one-body Density Matrix (OBDM) which is
equally valid at low and high densities.  The OBDM is calculated
using diffusion Monte Carlo methods and it is diagonalized to
obtain the ``natural" single particle orbitals and their
occupation, including the condensate fraction. At low Boson
density, $na^3 < 10^{-5}$, where $n = N/V$ and $a$ is the hard
core diameter, the condensate is localized at the center of the
trap. As $na^3$ increases, the condensate moves to the edges of
the trap.  At high density it is localized at the edges of the
trap. At $na^3 \leq 10^{-4}$ the Gross-Pitaevskii theory of the
condensate describes the whole system within $1\%$. At $na^3
\approx 10^{-3}$ corrections are $3\%$ to the GP energy but $30\%$
to the Bogoliubov prediction of the condensate depletion.  At
$na^3 \gtrsim 10^{-2}$, mean field theory fails. At $na^3 \gtrsim
0.1$, the Bosons behave more like a liquid $^4$He droplet than a
trapped Boson gas.
\end{abstract}
  %Bose Einstein condensation, 03.75.F

\maketitle

\section{Introduction}
Bose-Einstein condensation (BEC) has been a topic of fundamental
interest since it was first predicted by Einstein in
1924~\cite{Einstein1925a}. He showed that as a consequence of Bose
statistics~\cite{Bose1924a} a macroscopic fraction, $N_0/N$, of
the atoms in an ideal Bose gas can condense into a single quantum
state. London~\cite{London1938a,London1938b} postulated that
superfluidity in liquid $^4$He was a consequence of a transition
to BEC.  But liquid $^4$He is a strongly interacting, dense Bose
liquid and the connection between BEC in an ideal gas and
superfluidity was not at all clear \cite{Leggett2001a}. Similarly,
the many-body correlation effects induced by the inter-boson
interaction significantly reduce the condensate fraction even at
zero temperature~\cite{Ceperley1995a,Moroni1997a}. Modern direct
measurements~\cite{Glyde2000a} of BEC in liquid $^4$He find only
7.25\% of the liquid in the condensate at T = 0K.

The theoretical framework for treating an interacting Bose gas was
initiated in 1947 by Bogoliubov~\cite{Bogoliubov1947a}.  He
developed a perturbation expansion valid for low density and weak
interaction, $na^3 \ll 1$ (where n is the number density $N/V$ and
a is the hard core diameter of the Bosons), and small depletion of
the condensate, $(N-N_0)/N \ll 1$. About a decade later, Onsager
and Penrose~\cite{Onsager1956a} and L\"{o}wdin~\cite{Lowdin1955a}
formulated a definition of BEC in terms of the eigenvalues and
eigenvectors (natural orbitals) of the one-body density matrix
(OBDM).   An orbital with macroscopic occupation arising from
diagonalization of the OBDM is defined as the ``condensate
wave-function" or order parameter. This formulation allows direct
access to condensate properties at arbitrary density and does not
require a large condensate fraction.  The work in this paper is
based on the OBDM formulation of BEC which is rigorously valid for
a strongly interacting system~\cite{Leggett2001a}.

In 1995, experiments in weakly interacting dilute vapors of the
alkali atoms $^{87}$Rb, $^{23}$Na and $^7$Li in magnetic traps
provided direct evidence of a clear transition from a thermally
distributed cloud to macroscopic occupation of a single  quantum
state~\cite{Anderson1995a,Davis1995a,Bradley1995a}.  This long
awaited direct realization of BEC spawned a dramatic renewal of
interest in Bose systems and BEC. Since the densities in these
experiments were low (typical number densities were $10^{12}
 {\rm cm}^{-3}$ and $na^3 \approx 10^{-6}$ where $a$ is the s-wave
scattering length of the atoms), almost all of the theoretical
activity has focused on the weakly interacting gas limit and the
Gross-Pitaevskii (GP) equation~\cite{Dalfovo1999a}. The GP
equation provides an excellent mean field description of the
condensate at low density. This is a valid description of the
whole Bose gas in the dilute limit, $na^3 \ll 1$, where most of
the atoms are in the condensate. However, it is inaccurate for
strongly interacting systems in which the condensate fraction is
significantly depleted by quantum fluctuations.  Since the
experiments in 1995, only a handful of studies have attempted to
consider the properties of BEC beyond the dilute regime and the GP
description of the
condensate~\cite{Krauth1996a,Ziegler1997a,Timmermans1997a,Braaten1997a,Gruter1997a,
Pearson1998a,Holzmann1999a,Hutchinson1999a,Fabrocini1999a,Giorgini1999a,
Tanatar2000a,Galli2000a,Blume2001a,Fabrocini2001a,Banerjee2001a,Astrakharchik2002a,Cherny2002a,Draeger2002a,
Andersen2002a,Bao2002a,Blume2002a,Astrakharchik2002b}. Most of
this relatively small body of work rely on modified forms of the
GP equation which incorporate higher terms in the Bogoliubov
expansion that include effects of atoms outside the condensate
within a local density approximation. Unfortunately, the
condensate fraction and distribution in the trap calculated by
such methods become inaccurate as the density becomes greater than
$na^3 \gtrsim 10^{-3}$~\cite{Giorgini1999a}.

It has recently become possible to study Bose systems with tunable
interactions~\cite{Inouye1998a,Stenger1999c,Cornish2000a,Roberts2001b,Roberts2001c,Weber2003a}
for which densities of up to $na^3 \approx 1$ are obtainable.
Specifically, $^{85}$Rb at densities in the range $na^3 \approx
10^{-3}-10^{-1}$ has been investigated.  BEC in metastable helium
isotopes~\cite{Robert2001a,Westbrook2002a,DosSantos2001a} with
$na^3 \approx 10^{-4}$ and in atomic hydrogen~\cite{Fried1998a}
with $na^3 \approx 10^{-5}$, are also higher density Bose gases.
This makes the study of BEC and the role of interactions in
trapped Bose gases over a wide range of densities of direct
interest to experiment.

\begin{figure}
\includegraphics[width=.49\textwidth]{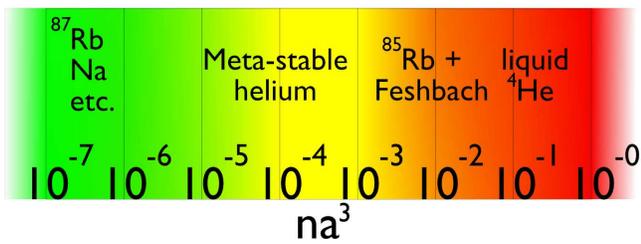}
\caption{Range of system densities considered in this work
expressed in terms of $na^3 \equiv {N a^3}/{V}$, the ratio of the
volume occupied by $N$ hard core particles with diameter $a$ to
the total volume of the system, $V$.} \label{density-range-fig}
\end{figure}

The chief purpose of this work is to go beyond the dilute limit,
to test the limits of the GP equation and related mean field
approximations and to explore the zero temperature properties of
trapped hard core Bosons as $na^3$ increases from the dilute limit
to the dense regime corresponding to liquid $^4$He, and beyond.
The range of densities investigated here is displayed in
Fig.~\ref{density-range-fig}. We increase the density by
increasing both $N$ and the hard core diameter, $a$, up to the
value $na^3 \simeq 0.21$ which describes liquid $^4$He at SVP when
the $^4$He atoms are represented by hard spheres of diameter $a =
2.203$\AA~\cite{Kalos1974a}.  The ground state energy, $E$, the
total density distribution, and the OBDM are evaluated using
diffusion Monte Carlo (DMC) methods.

Specifically, we compare the ground state energy of the whole
trapped gas calculated using DMC, $E_{DMC}$, with the usual energy
of the condensate calculated using the GP equation, $E_{GP}$.  As
density increases, $E_{DMC}$ and $E_{GP}$ begin to differ. For
example, at $na^3=10^{-3}$, we find $(E_{DMC}-E_{GP})/E_{GP} =
3\%$.  Modified GP equations provide a mean field description of
the atoms above the condensate.  The dependence of
$E_{DMC}-E_{GP}$ on the number of trapped Bosons, $N$, and on the
scattering length, $a$, follows the predictions of the Modified GP
equation remarkably well up to high densities, $na^3 \approx
5\times 10^{-2}$.  This suggests that the difference
$E_{DMC}-E_{VMC}$ can be attributed to the atoms above the
condensate.  However, the energy is not as sensitive to
approximations as some other properties.

We compare the condensate fraction obtained using the rigorous
OBDM-DMC method with predictions of the Bogoliubov theory.  The
two agree within $1\%$ for $na^3 \lesssim 10^{-4}$.  At higher
densities, the Bogoliubov theory significantly underestimates the
depletion of the condensate, by $25\%$ at $na^3 \approx 2\times
10^{-2}$. We evaluate the condensate density distribution in the
trap.  At low density, the condensate is localized at the center
of the trap as usually found \cite{Dalfovo1999a}.  At higher
density ($na^3 \approx 10^{-2}$), the condensate is spaced over
several trap lengths and the condensate and total density have
similar distributions.  Also, at higher densities ($na^3 \gtrsim
2\times 10^{-2}$), oscillations in the total density distribution
appear which are not found in mean field theories. There are no
corresponding oscillations in the condensate density distribution.
At high density ($na^3 \gtrsim 0.10$), the condensate is localized
at the edges of the trap (large $r/a$) where the total boson
density is low.     At high density, the trapped Bosons resemble
liquid $^4$He droplets \cite{Dalfovo2001a,DuBois2001a,Chin1993a}.

We also compare the present DMC results with our earlier
variational Monte Carlo (VMC) values~\cite{DuBois2001a}.  We find
that the VMC and DMC energies agree well at all densities and that
the ground state energy is not very sensitive to the trial
variational wave-function.  However, the OBDM and the condensate
fraction is very sensitive to the trial wave function at higher
densities.  An accurate initial trial function is needed to get
reliable condensate fractions even in the DMC formulation.

Monte Carlo methods are usually applied to dense systems such as
liquid and solid $^{4}$He~\cite{Ceperley1995a,Moroni1997a}.
Recently Giorgini et al. have evaluated the energy and condensate
fraction of the uniform Bose gas over a wide density range,
$10^{-6} \leq na^3 \leq 10^{-3}$ \cite{Giorgini1999a}.  Gr{\"u}ter
et al.~\cite{Gruter1997a}, have evaluated the critical
temperature, T$_c$, for BEC in a Bose gas using path integral
Monte Carlo (PIMC) methods. They find T$_c$ is increased above the
ideal Bose gas value by interaction in the dilute range. This
increase is observed in dilute concentrations of $^4$He in
Vycor~\cite{Reppy2000a}.  At liquid $^4$He densities, T$_c$ is
decreased by interaction \cite{Wilks1967a,Pollock1992a}.

Krauth~\cite{Krauth1996a} first applied QMC to BEC in a trap using
PIMC methods.  For $10,000$ hard sphere Bosons in a spherical trap
with a ratio of hard core diameter to trap length, $a/a_{ho} =
4.3\times10^{-3}$ ($na^3 \approx 10^{-4}$), he found that
condensate was concentrated at the center of the trap while the
uncondensed atoms were spread over a wide range and well described
by a classical Bose gas.  Holzmann et al. \cite{Holzmann1999a}
made a direct comparison of PIMC and Hartree-Fock calculations for
a dilute gas of hard spheres in a trap with $a/a_{ho} = 4.3\times
10^{-3}$.  For temperatures near T$_c$, they found $N_0$ was
greater in PIMC.  The increase in $N_0$ with exact representation
of the interaction effects is consistent with the corresponding
increase in $T_{c}$ with interaction in the uniform Bose gas.

Recently, QMC methods have been successfully applied to the study
of highly inhomogeneous Bose systems.  Astrakharchik et al. used
DMC to study BEC and superfluidity in a Bose gas with disorder at
zero temperature \cite{Astrakharchik2002a}. They find an
intriguing decoupling of the superfluid and condensate fractions
for strong disorder. Studies of superfluid $^4$He with a free
surface~\cite{Galli2000a,Draeger2002a} found the local condensate
fraction peaks ($n_0 \approx 0.95\%$~\cite{Draeger2002a}) in the
dilute region just inside the liquid-vacuum interface.
Blume~\cite{Blume2002a} and Astrakharchik and
Giorgini~\cite{Astrakharchik2002b} have examined the transition
from the three-dimensional to the quasi one-dimensional regime for
Bosons in highly elongated cigar-shaped traps.  They confirm that
the Bose gas undergoes ``fermionization" in the quasi 1-D regime.

In section \ref{method section}, we describe the theoretical
framework and computational methods used. Section \ref{results
section} contains the present results.  In section \ref{discussion
section}, the chief results are reviewed and discussed.

\section{Methods}
\label{method section} We consider $N$ Bosons of mass $m$ confined
in an external trapping potential, $V_{ext}({\bf r})$, and
interacting via a two-body potential $V_{int}({\bf r}_1,{\bf
r}_2)$. The Hamiltonian for this system is:
\begin{equation}
    H = \sum_i^N \left(
        \frac{-\hbar^2}{2m}
        { \nabla }_{i}^2 +
        V_{ext}({\bf{r}}_i)\right)  +
        \sum_{i<j}^{N} V_{int}({\bf{r}}_i,{\bf{r}}_j).
\end{equation}
Here,
\begin{equation}
V_{ext}({\bf r}) = \frac{1}{2}m\omega_{ho}^2r^2,
\label{trap_potential}
\end{equation}
where $\omega_{ho}^2$ is the characteristic trap frequency.
Interactions are modelled by a hard sphere potential,
\begin{equation}
V_{int}(r) =  \Bigg\{
\begin{array}{ll}
        \infty & {r} \leq {a}\\
        0 & {r} > {a}.
\end{array}
\label{hard_spheres}
\end{equation}

Introducing lengths in units of the characteristic trap length
$a_{ho} = (\hbar/m\omega_{ho})^{1/2}$, $r \rightarrow r/a_{ho}$,
and energies in units of $\hbar\omega_{ho}$ as in
\cite{Dalfovo1999a}, the many-body Hamiltonian is:
\begin{equation}
H = \sum_i^N\frac{1}{2}(-\nabla _{i}^2 + r_i^2) +
    \sum_{i<j} V_{int}(|{\bf r}_i-{\bf r}_j|).
\end{equation}

\subsection{Diffusion Monte Carlo implementation}
Diffusion Monte Carlo is a method for finding the exact properties
of the quantum mechanical ground-state of a many-body system to
within arbitrary precision -- see, for example, Ref.
\cite{Reynolds1982a}. The starting point for this method is the
time dependent Schroedinger equation in imaginary time:
\begin{equation}
\label{tds2} [-\frac{\hbar^2}{2m}{\nabla}^2 + V({\bf R})
-E_T]\Psi({\bf R},t) = -\hbar\frac{\partial {\Psi({\bf
R},t)}}{\partial {t}}.
\end{equation}
The time dependent component of $\Psi({\bf R},t)$, $Q_i(t)$, is
$Q_i(t)=exp[-(E_i-E_T) t/\hbar]$.  $E_T$ is an adjustable target
energy. In the $t \rightarrow \infty$ limit, the steady state
solution of (\ref{tds2}) is the ground state $\Phi_0({\bf R})$.

The term diffusion Monte Carlo comes from the resemblance of
(\ref{tds2}) to the classic diffusion equation:
\begin{equation}
\label{deqn} D{\nabla}^2\rho({\bf R},t) = \frac{\partial
{\rho({\bf R},t)}}{\partial {t}}.
\end{equation}
This equation can be simulated by a Monte Carlo random walk in
configuration space.  Treating the $[V({\bf R}) -E_T]\Psi({\bf
R},t)$ component of (\ref{tds2}) alone results in a rate equation
of the form
\begin{equation}
\label{reqn} v({\bf R})\rho({\bf R},t) = -\frac{\partial
{\rho({\bf R},t)}}{\partial {t}}.
\end{equation}
This component represents a branching process in which the growth
or decay of a population is proportional to its density. In the
present implementation the diffusion and branching processes are
combined to simulate (\ref{tds2}) and obtain the zero temperature
ground state of the time independent Schrodinger equation.

A simple application of (\ref{tds2}) above results in a branching
rate which is proportional to the potential energy $V({\bf R})
-E_T$.  This means that large fluctuations in the potential,
$V({\bf R})$, will cause correspondingly large fluctuations in the
population of walkers.  Dramatic fluctuations in the number of
walkers can result in large inefficiencies when treating realistic
many-body systems.   The solution to this problem was first
presented by Kalos et al. \cite{Kalos1974a}. In this method, a
trial function is introduced to {\em guide} the metropolis walk to
regions of higher probability and lower potential energy resulting
in lower fluctuations in the population of walkers.  The
wave-function in (\ref{tds2}) is replaced by a product of the true
ground state, $\Psi({\bf R},t)$, and a guiding function
$\Psi_T({\bf R})$,
\begin{equation}
\Psi({\bf R},t) \rightarrow \Psi({\bf R},t)\Psi_T({\bf R}).
\end{equation}
While use of a guiding function is necessary for the efficient
application of the DMC method, it can introduce a bias into the
calculation of observables which do not commute with the
Hamiltonian unless corrective measures are taken -- such as the
application of ``forward walking"
\cite{Kalos1974a,Casulleras1995a}.

We evaluate the expectation value, $\langle \Psi| \mathcal{O}
|\Psi \rangle$, of an operator $\mathcal{O}$, using QMC. In
integral form the expectation value is
\begin{equation}
\langle \Psi| \mathcal{O} |\Psi \rangle = \int d{\bf R}~
\Psi^{*}({\bf R}) \mathcal{O}({\bf R}) \Psi({\bf
R}).\label{operator_integral}
\end{equation}
To evaluate this expression using QMC, (\ref{operator_integral})
is recast as
\begin{equation}
 \langle \Psi| \mathcal{O} |\Psi \rangle = \int d{\bf R}~
 |\Psi({\bf R})|^2
 \Big[
    \frac
        {\mathcal{O}({\bf R}) \Psi({\bf R})}
        {\Psi({\bf R})}
  \Big].
  \label{local expression}
\end{equation}
The result of a QMC calculation is a set of configurations $\{{\bf
R_1},...,{\bf R_M}\}$ sampled from $|\Psi|^2$.  Using these
configurations we may estimate $\langle \Psi | \mathcal{O} | \Psi
\rangle$ as
\begin{equation}
\langle \Psi | \mathcal{O} | \Psi \rangle \approx
\frac{1}{M}\sum_{i=1}^M
 \frac{\mathcal{O}({\bf R_i})\Psi({\bf R_i})}{\Psi({\bf R_i})}.
\end{equation}
This estimate becomes exact as $M \rightarrow \infty$.

\subsection{The OBDM and natural orbitals}
A goal in this work is to describe BEC in systems with
interactions. To do this we require a definition of the condensate
single particle state. Following Penrose and Onsager, L\"owdin and
others \cite{Onsager1956a,Lowdin1955a}, we take the one-body
density matrix (OBDM) as the fundamental quantity for an
interacting system and define the natural single particle orbitals
(NO) in terms of the OBDM.  The OBDM is \cite{Baym1976a}
\begin{equation}
 \rho({\bf r'},{\bf r}) = \langle \hat{\Psi}^{\dagger}({\bf r}'),\hat{\Psi}({\bf r}) \rangle,
\end{equation}
where $\hat{\Psi}({\bf r})$ is the field operator that annihilates
a single particle at the point ${\bf r}$ in the system. To define
the NO, we introduce a set of single particle states having wave
functions $\phi_i({\bf r})$ and expand $\hat{\Psi}({\bf r})$ in
terms of these states and the operators $\hat{a}_i$ which
annihilate a particle from $|i\rangle$,
\begin{equation}
\hat{\Psi} =\sum_i\phi_i({\bf r})\hat{a}_i.
\end{equation}
Requiring that the $\hat{a}_i$ satisfy the usual commutation
$([\hat{a}_i^{\dagger},\hat{a}_j] = \delta_{ij})$ and number
relations $(\langle \hat{a}_i^{\dagger}\hat{a}_j
\rangle=N_i\delta_{ij})$, we have
\begin{equation}
\begin{array}{r}
\rho({\bf r},{\bf r}') =
        \mathop{\sum}_{ij} \phi_j^{*}({\bf r}')\phi_i({\bf r})N_i\delta_{ij}\\
        =\mathop{\sum}_{ij} \phi_j^{*}({\bf r})\phi_i({\bf r}')N_i\delta_{ij}
\end{array}
\label{OBDM-defined}
\end{equation}
This may be taken as the defining relation of the NO, $\phi_i({\bf
r})$. Specifically, we have from (\ref{OBDM-defined}),
\begin{equation}
        \int{
                d{\bf r}d{\bf r}'
        \phi_i^{*}({\bf r})
        \rho({\bf r},{\bf r}')
        \phi_j({\bf r}')
        = N_i\delta_{ij}
        },\label{obdm_occupation}
\end{equation}
so that the NO may be obtained by diagonalizing the OBDM.  The
eigenvectors are the NO and the eigenvalues are the occupation,
$N_i$, of the orbitals.  In principle any orbital which satisfies
$N_i  >> 1$ may be considered a macroscopically occupied
pseudo-particle state -- i.e. the equivalent of a Bose-Einstein
condensate. A Bose system with more than one macroscopically
occupied state would represent a fragmented condensate
\cite{Leggett2001a}. In the systems studied in this work, only a
single condensate orbital was found to have macroscopic
occupation. The condensate is therefore the orbital having the
highest occupation, denoted $\phi_0(r)$, and the condensate
fraction is $n_0=N_0/N$.

The relations (\ref{OBDM-defined}) and (\ref{obdm_occupation})
involve the vector ${\bf r}$ and ${\bf r}'$ and cannot be solved
directly as matrix equations. To obtain matrix equations, we
restrict ourselves to spherical traps and seek equations for the
radial component of the NO as in ref. \cite{DuBois2001a}.  In this
approach, the OBDM is expanded in Legendre Polynomials,
$P_l(\hat{\bf r}_{_1}\cdot\hat{\bf r}'_{_1})$, and evaluated using
the QMC ground state, $\Psi_0$, as
\begin{equation}
\rho_{_l}(r_{_1},r_{_1}') =
        \int{
                d\Omega_1d{\bf r}_{_2}..d{\bf r}_{_N}
                \Psi_0^{*}({\bf r}_{_1}..{\bf r}_{_N})
                P_l(\hat{\bf r}_{_1}\cdot\hat{\bf r}'_{_1})
                \Psi_0({\bf r}'_{_1}..{\bf r}_{_N})
        }.
        \label{obdm_spherical_form}
\end{equation}

\subsection{QMC Evaluation of $\rho_l(r,r')$}

In QMC we evaluate (\ref{obdm_spherical_form}) in a form similar
to (\ref{local expression}) giving
\begin{equation}
\begin{array}{l}
\rho_l(r_1,r_1') \approx \\
    \frac{1}{4\pi\epsilon}\int_{r_1-\epsilon/2}^{r_1+\epsilon/2}{
        d r_1
        \int d\Omega_1d{\bf \tilde{R}}
        |\Psi({\bf r}_1,{\bf \tilde{R}})|^2
        \Big[
            \frac
            {P_l(\hat{r}_{_1}\cdot\hat{r}'_{_1}) \Psi({\bf r}_1',{\bf \tilde{R}})}
            {\Psi({\bf r}_1,{\bf \tilde{R}})}
        \Big]
    },
\end{array}\label{rho_l using QMC}
\end{equation}
where ${\bf \tilde{R}} \equiv ({\bf r}_2,..,{\bf r}_N)$
 and
$\epsilon$ is the width of the grid elements upon which
$\rho_l(r_1,r_1')$ is being evaluated.
 Because the
systems we are evaluating are spherically symmetric, the direction
of ${\bf r}'$ is arbitrary. We may take advantage of this fact to
reduce the statistical uncertainty in estimates of $\rho_l(r,r')$
by evaluating (\ref{rho_l using QMC}) for several different
directions of ${\bf r}'$ and taking the average result. In
addition, since we are dealing with identical Bosons, the OBDM
does not depend on the particle being evaluated so
$\rho_l(r_1,r_1') = \rho_l(r_i,r_i')$. This allows us to take the
average, $\rho_l(r_1,r_1') = 1/N \sum_i \rho_l(r_i,r_i')$.

\subsection{Diagonalization and error estimation} Using the
method described above, the OBDM is evaluated on a grid of values
of $r = i\epsilon$ and $r' = j\epsilon$ where i and j are integers
in the range $0 \leq i,j \leq Q$ (where $Q$ is a maximum cutoff).
We may then construct the discreet matrix,
$[i\epsilon\rho_{_l}(i\epsilon,j\epsilon)j\epsilon]$, which is
readily diagonalized by standard matrix diagonalization methods.

Replacing the continuous matrix $\rho_l(r,r')$ with the discreet
matrix $[i\epsilon\rho_{_l}(i\epsilon,j\epsilon)j\epsilon]$ is a
potential source of systematic error.  To avoid this problem, we
evaluated each system with decreasing values of the grid spacing,
$\epsilon$, such that $\epsilon_{q+1} = \epsilon_{q}/2$. The
largest value of $\epsilon$ for which no significant change in the
calculated orbitals and occupation numbers occurred between
$\epsilon_q$ and $\epsilon_{q+1}$ was then used to determine the
condensate properties for that system.

A second potential source of error arises in treating
$\rho_l(r,r')$ (which is an infinite matrix) as a finite matrix.
Since the trapped systems are spatially finite, the probability of
finding a particle beyond the average radius, $R$, of the cloud
goes to zero very quickly.  For the same reason, $\rho_l(r,r')
\approx 0$ when either $r > R$ or $r' > R$. It is therefore, safe
to treat $\rho_l(r,r')$ as a finite matrix.  As a brute force test
of this assertion, we evaluated several systems with increasingly
large cutoff values.  We found no significant change in condensate
properties calculated from an OBDM where $r,r' \leq R$ and $r,r'
\leq 2R$.

The statistical error associated with a given orbital and its
occupation are obtained as follows. When the initial OBDM,
$\rho^0$, is calculated the variance associated with each matrix
element in $\rho^0$ is obtained.  The original $\rho^0$ is assumed
to represent a randomly sampled event from a gaussian error
distribution surrounding the true OBDM. Based on this assertion, a
set of M new OBDM's, $\{\tilde{\rho}^1..\tilde{\rho}^M\}$, are
then generated by allowing each matrix element to randomly vary
according to its statistical error.  Each of the new OBDM,
$\tilde{\rho}^q$, are diagonalized to obtain their corresponding
eigenvalues, $\tilde{n}_i^q$ and eigenvectors $\tilde{\phi}_i^q$.
An average occupation, $\overline{n}_i = 1/M
\sum^M_q{\tilde{n}_i^q}$, and orbital, $\overline{\phi}_i =
\sum^M_q\tilde{\phi}_i^q$, are then obtained.  The variance of
these averages is then used as an estimate of the statistical
error of the orbitals and occupation numbers of $\rho^0$.

\section{Results}
\label{results section}
\subsection{DMC Energy}
\begin{figure}
\includegraphics[width=.49\textwidth,bb = 109 329 509 649,clip]{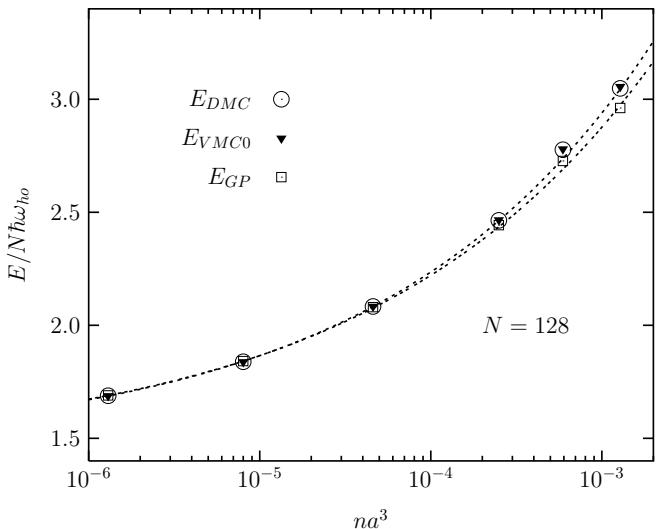}
\caption{Diffusion Monte Carlo, $E_{DMC}$, variational Monte
Carlo, $E_{VMC0}$, and Gross-Pitaevskii, $E_{GP}$, energies for
trapped hard sphere Bosons as a function of maximum density,
$na^3$, in the trap. Density is varied by changing scattering
length $a$, $4.3 \times 10^{-3}< a/a_{ho} < 0.14$ where $a_{ho}$
is the trap length.  At higher densities $E_{DMC}$ clearly lies
above $E_{GP}$, $3\%$ at $na^3 = 10^{-3}$. $E_{VMC}$ and $E_{DMC}$
differ by $0.3\%$ at $na^3 = 10^{-3}$.}
\label{E_DMC-vs-na3-dilute}
\end{figure}

Figure \ref{E_DMC-vs-na3-dilute} shows the energy per particle
calculated by diffusion Monte Carlo, $E_{DMC}$, by variational
Monte Carlo (using the simple trial function
of~\cite{DuBois2001a}), $E_{VMC0}$, and using the Gross-Pitaevskii
equation, $E_{GP}$, of trapped hard core Bosons as a function of
maximum density, $na^3$, in the trap.  In the dilute regime, $na^3
\lesssim 10^{-4}$, $E_{DMC}$, $E_{VMC0}$, and $E_{GP}$ are nearly
indistinguishable. The difference in energy at $na^3 = 5 \times
10^{-5}$ is, for example $10^{-3}\hbar\omega_{ho}$ which is within
the error bars of the QMC calculations. At higher densities, the
DMC energy lies above the GP result by $3\%$ at $na^3 = 10^{-3}$.
$E_{VMC0}$ agrees well with the DMC results with a difference of
only $0.3\%$ at $na^3 = 10^{-3}$.

\begin{figure}
\includegraphics[width=.49\textwidth,bb= 86 401 473 721,clip]{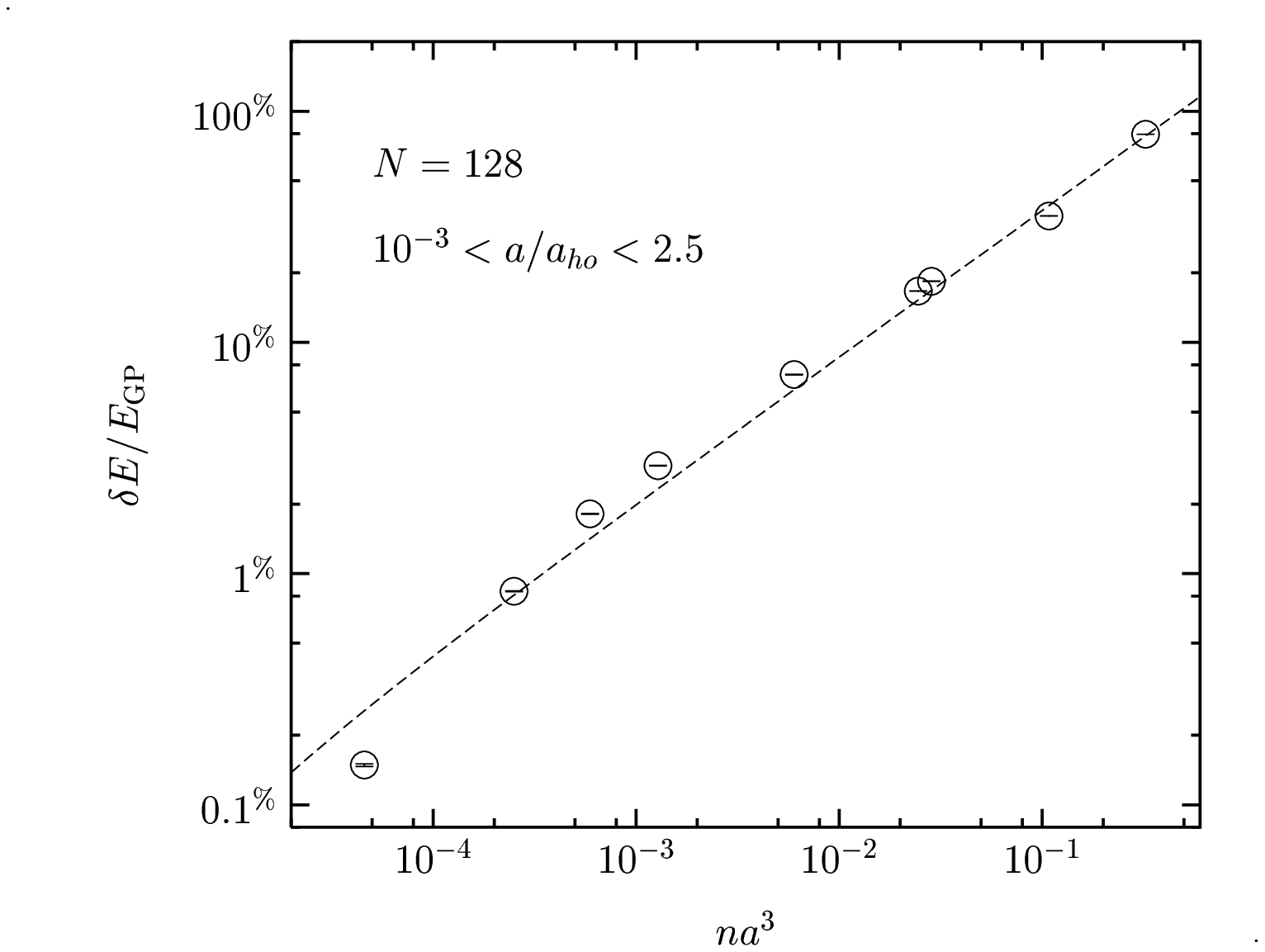}
\caption{Percent difference between diffusion Monte Carlo,
$E_{DMC}$, and Gross-Pitaevskii, $E_{GP}$, energies for hard core
Bosons in a spherically symmetric harmonic trap as a function of
maximum density $na^3$ in the trap. The \% difference between DMC
and GP energies is well described by $\delta E / E_{GP} \propto
(na^3)^{2/3}$ (dashed line).} \label{E_DMC-E_GP-vs-na3.eps}
\end{figure}

Figure \ref{E_DMC-E_GP-vs-na3.eps} shows the percent difference
between $E_{DMC}$ and $E_{GP}$, $\delta E / E_{GP} =
({E_{DMC}-E_{GP}})/{E_{GP}}$ for $N=128$ hard sphere Bosons in a
spherically symmetric harmonic trap at higher densities, $na^3$.
Here, and throughout this paper, GP energies are calculated using
a self interaction term proportional to $(N-1)a/a_{ho}$. GP
results using $Na/a_{ho}$ significantly overestimate the energy
for small $N$.
 The difference between DMC and GP
energies is well described by $\delta E / E_{GP} \propto
(na^3)^{2/3}$. This dependence holds even up to trap densities of
$na^3 \approx 0.32$, well above the density of liquid helium
($na^3 \approx 0.21$). At this density, $E_{GP}$ and $E_{DMC}$
differ by as much as $80\%$.

\begin{figure}
\centerline{\includegraphics[width=.49\textwidth,bb= 116 402 510
723, clip]{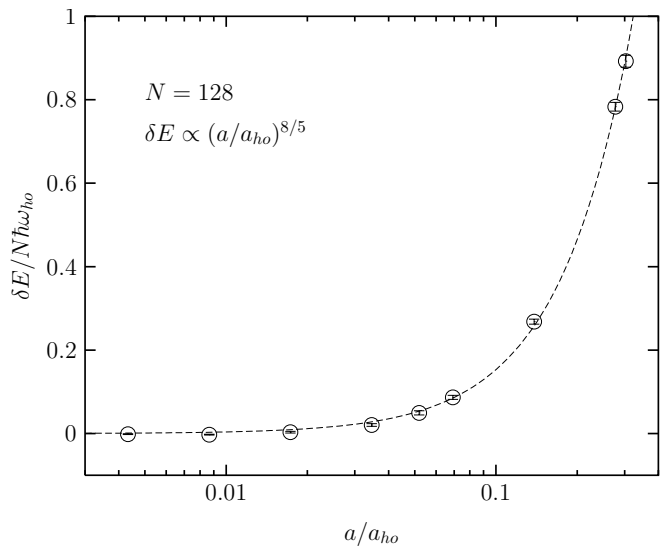}} \caption{Dependence of
$\delta E = ({E_{DMC} -E_{GP}})$ on the ratio of the scattering
length, $a$, to the trap length, $a_{ho} =
(\hbar/m\omega_{ho})^{1/2}$, for $N=128$ Bosons in a spherically
symmetric harmonic trap.  The dashed line shows $\delta E/E_{GP}
\propto (a/a_{ho})^{8/5}$.} \label{E_DMC-E_GP-vs-a.eps}
\end{figure}

In Fig. \ref{E_DMC-E_GP-vs-a.eps}, the dependence of $\delta
E/E_{GP}$ on the scattering length, $a$, for $N=128$ Bosons in a
spherically symmetric harmonic trap is shown.  The figure shows
good agreement with $\delta E \propto (a/a_{ho})^{8/5}$. This is
precisely the power law relation predicted by the first-order
correction to the Gross-Pitaevskii energy which takes into account
particles above the condensate, denoted the modified
Gross-Pitaevskii equation (MGP) energy \cite{Dalfovo1999a}.

\begin{figure}[ht!]
\centerline{\includegraphics[width=.49\textwidth,bb=157 308 552
631,clip ]{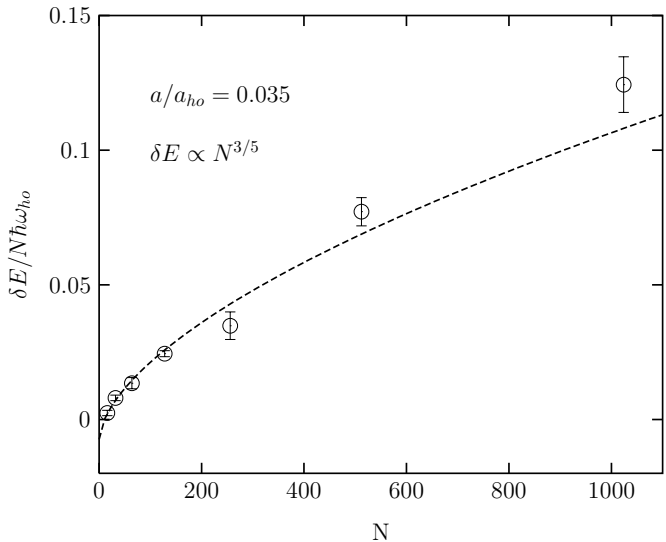}} \caption{Dependence of $\delta E
= E_{DMC} -E_{GP}$ on the number of particles,\ $N$, in units of
$\hbar \omega_{ho}$ where the ratio of the scattering length to
the characteristic length of the trap is $a/a_{ho} = 0.03464$. The
dashed line is $\delta E \propto N^{3/5}$.} \label{dE-vs-N.eps}
\end{figure}

Figure \ref{dE-vs-N.eps} shows the dependence of $\delta E =
E_{DMC}-E_{GP}$ on the number of particles, $N$, in the trap. In
this plot, the ratio of the scattering length to the
characteristic length of the trap is $a/a_{ho} = 8 \times
a_{Rb}/a_{ho} = 0.03464$. The resulting range of densities at the
center of the trap lie between $na^3 \approx 8\times 10^{-5}$ for
$N=16$ and $na^3 \approx 6 \times 10^{-4}$ for $N=1024$. The DMC
energy is approximately 2\% higher than the GP energy when $N =
1024$. The dashed line is a least squares fit of $\delta E$ to a
function of the form $q(N) = q_0 + q_1N^{3/5}$. The relation
$\delta E(N) \propto N^{3/5}$ is again consistent with the result
obtained from the modified Gross-Pitaevskii equation.

\begin{figure}[ht!]
\centerline{\includegraphics[width=.49\textwidth,bb=158 304 563
637,clip]{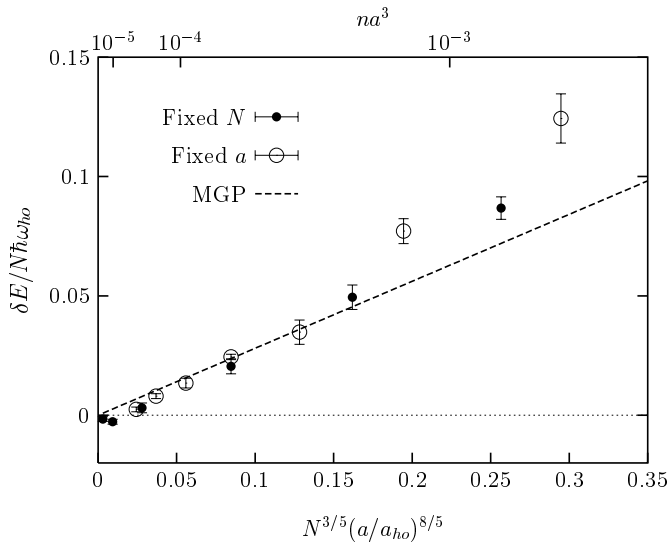}} \caption{$\delta E = ({E_{DMC}
-E_{GP}})$ as a function of $N^{3/5}(a/a_{ho})^{8/5}$ for fixed
number of particles, $N=128$, (filled circles) and fixed
scattering length, $a/a_{ho} = 8 \times a_{Rb}$, (open circles)
along with MGP prediction (heavy dashed line). Values of the
maximum trap density, $na^3$, for the fixed $N$ case are shown on
the top axis. } \label{dE-vs-Na.eps}
\end{figure}

The MGP expression for the ground state energy provides the first
correction to the GP energy arising from contributions of the
noncondensate.  If this correction is relevant across the entire
range of systems considered, combining the results for the
dependence of $\delta E$ on $N$ and $a/a_{ho}$ as presented in
figures~\ref{E_DMC-E_GP-vs-a.eps} and \ref{dE-vs-N.eps} should
provide a single coefficient, $\xi$, such that
\begin{equation}
    \delta E = \xi N^{3/5}(a/a_{ho})^{8/5}. \label{dE-LDA}
\end{equation}
MGP predicts $\xi = 5(15)^{3/5}/(64\sqrt{2}) \approx 0.28$. In
Fig.~\ref{dE-vs-Na.eps}, the fixed $a$ and fixed $N$ results are
shown together along with the MGP prediction for the first order
contribution to the ground state energy of atoms depleted from the
condensate. The figure demonstrates that for systems with $na^3
\lesssim 5 \times 10^{-4}$, MGP provides a good description of the
DMC corrections to the GP energy.  At higher densities, while the
fixed $a$ and fixed $N$ results are separately well described by
$(a/a_{ho})^{8/5}$ and $N^{3/5}$ respectively, they do not share a
common coefficient $\xi$.  This suggests that at higher densities
corrections to the condensate energy have a more complicated
dependance on $N$ and $a$ than~(\ref{dE-LDA}).

\subsection{Range of validity of VMC0 results}
\begin{figure}
\includegraphics[width=.49\textwidth,bb=93 390 473 712,clip]{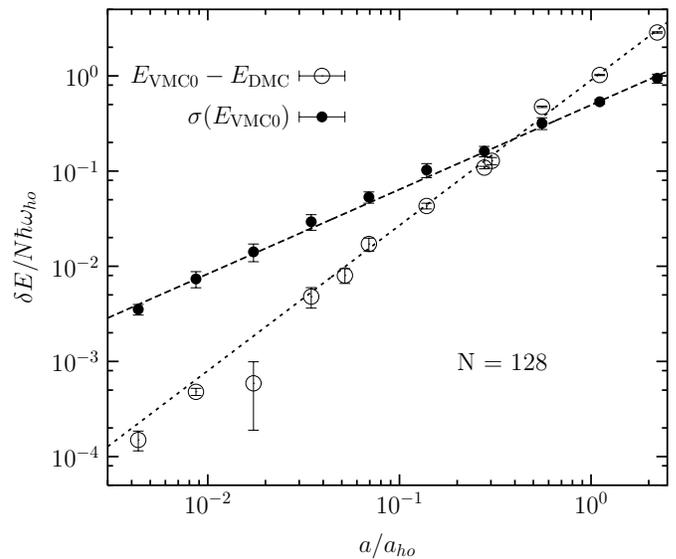}
\caption{Difference between DMC and VMC0 energies
$(E_{VMC0}-E_{DMC})$ compared with the variance of the VMC0
calculation, $\sigma(E_{VMC0})$, as a function of the ratio of the
hard sphere diameter to the trap length, $a/a_{ho}$.
}\label{E_DMC-E_VMC0-vs-sigmasq_VMC0.eps}
\end{figure}

To investigate the range of validity of the VMC0 trial function we
evaluated the variance of the Hamiltonian. If the trial function
is an exact representation of an eigenstate of the Hamiltonian the
variance is zero.   Figure \ref{E_DMC-E_VMC0-vs-sigmasq_VMC0.eps}
provides a comparison of the difference between DMC and VMC0
results for the energy per boson, $(E_{VMC0}-E_{DMC})$, and the
variance of the energy per boson, $\sigma(E_{VMC0})$, as a
function of the ratio of the hard sphere diameter to the trap
length, $a/a_{ho}$. Results are for $N=128$ hard core Bosons in a
spherically symmetric trap.  Up to a value of $a/a_{ho}
\approx~0.3$, the DMC and VMC0 energies agree to within the
variance of $E_{VMC0}$.  The maximum density of the trapped Bosons
for this ``critical" value of $a/a_{ho} = 0.3$ is $na^3 \approx 3
\times 10^{-2}$.  This indicates that for systems with $na^3
\lesssim 10^{-2}$ the VMC0 trial function not only provides a
valid upper bound on the energy but a valid lower bound as well.

\subsection{Spatial distribution of trapped Bosons}
\begin{figure}
\includegraphics[width=.49\textwidth,bb=157 304 553 632,clip]{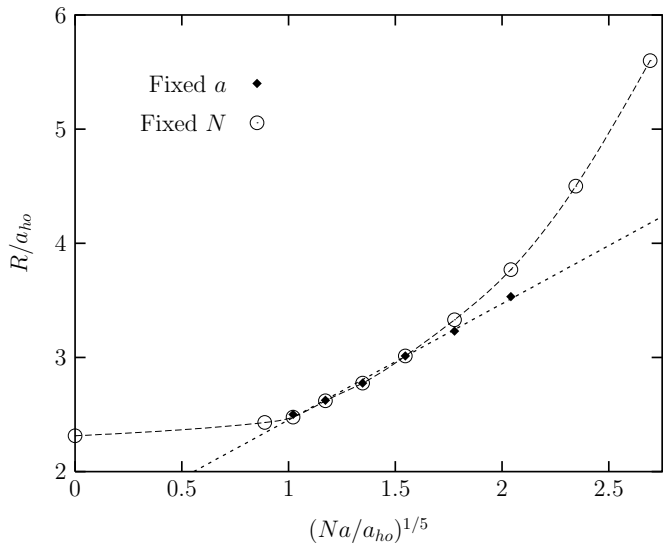}
\caption{QMC values of the width, $R$, of the ground state density
of hard-core Bosons in a harmonic trap verses $({
N}a/a_{ho})^{1/5}$ where ${ N}$ is the number of particles and $a$
is the hard core diameter. Diamonds show the dependence when ${
N}$ is fixed (${ N} = 128$) and $a$ is varied, $4.33 \times
10^{-3} < a/a_{ho} < 1.11$.
 Circles show the dependence when $a$ is fixed ($a/a_{ho} = 8\times a_{Rb}/a_{ho}
\approx 0.035$) and ${ N}$ is varied, $32 \leq { N} \leq 1024$.
The short-dashed and long-dashed lines are linear and spline fits
to the fixed $a$ and fixed $N$ data respectively.
}\label{width-vs-Na.eps}
\end{figure}

The spatial distribution of trapped Bosons is a property which is
accessible to experimental observation.  The first observations of
BEC used the difference between a classical Boltzmann distribution
and a condensate distribution as evidence for the existence of BEC
\cite{Anderson1995a,Davis1995a,Bradley1995a}. Spatial resolution
in most observations to date is, however, not very high (typically
only ${\mathcal O}(10^{-1})$ times the size of the condensate
itself) \cite{Anderson1995a,Roberts2001c}.  In this section, we
compare the present QMC results for the density of the many-body
ground state, $n(r)$, with predictions of mean-field theory for
the spatial distribution of the condensate, $n_0(r)$.  While this
comparison is not always strictly correct, since depletion of the
condensate means $n(r) \neq n_0(r)$, what is actually observed in
experiments is the ``total" density which includes condensate and
non-condensate atoms alike.  The condensate distribution and
``total" density have been treated as identical in the analysis of
experimental results \cite{Cornish2000a}.  For this reason, we
will compare $n(r)$ and mean field results for $n_0(r)$ as if they
are indeed measurements of the same physical quantity.

\subsubsection{The ``width" of a trapped cloud of Bosons}
\label{DMC-vs-TF-Width}
 The radius of the condensate as predicted by the
Gross-Pitaevskii equation in the Thomas-Fermi limit ($Na >> 1$,
$a/a_{ho} << 1$) is \cite{Dalfovo1999a}
\begin{equation} R_{TF} =
a_{ho}(15 N \frac{a}{a_{ho}})^{1/5}. \label{TF-radius}
\end{equation}
We have defined the radius of the ground state, $R_{QMC}$ by
setting a cut-off value of the QMC number density, $n(r)$, so that
$n(R_{QMC}) = 10^{-5}$. Figure \ref{width-vs-Na.eps}. shows QMC
results for the dependence of the width, $R/a_{ho}$, of the ground
state density of $N$ hard-core Bosons on the product
$(Na/a_{ho})^{1/5}$. In the figure, diamonds show the dependence
when the number of particles is fixed, $N = 128$, and the hard
core diameter, $a$, is varied, $4.33 \times 10^{-3} < a/a_{ho} <
1.11$. The dashed line is a spline fit to the fixed $N=128$ data
to guide the eye. Circles show the dependence when the hard-core
diameter is fixed, $a/a_{ho} = 0.035$, and $N$ is varied, $32 \leq
N \leq 1024$. The dashed line is a linear least squares fit to the
fixed-$a$ data with slope $\approx 0.52$. In the region
$1~\lesssim~(Na/a_{ho})^{1/5}~\lesssim~1.75$, both fixed-$a$ and
fixed-$N$ results have a linear dependence on $(Na/a_{ho})^{1/5}$
with the same slope.  In the region where the dependence on
$(Na/a_{ho})^{1/5}$ holds, the maximum density of the trapped
Bosons ranges from $10^{-6}~\lesssim~na^3~\lesssim~5 \times
10^{-3}$. For small values of $(Na/a_{ho})$, the
$(Na/a_{ho})^{1/5}$ dependence is not expected to hold since even
a single particle non-interacting system has a finite width. The
width of the many-body ground state is no longer linearly
dependent on $(a/a_{ho})^{1/5}$ for values of $(Na/a_{ho})^{1/5}
> 1.75$.  In this regime, the maximum trap density is
$na^3~\gtrsim 5 \times 10^{-3}$ and $a/a_{ho}~\gtrsim~0.1$.  The
present DMC results indicate that for $a/a_{ho}~\gtrsim~0.1$, the
width of the many-body ground state depends on $a/a_{ho}$ as $R
\propto (a/a_{ho})^{2/3}$ rather than $(a/a_{ho})^{1/5}$. Linear
dependence on $N^{1/5}$ continues to hold up to the highest number
of particles considered ($N=1024$).

\subsubsection{The Total Density Profile}

\begin{figure}
 \centerline{\includegraphics[width =.49\textwidth]{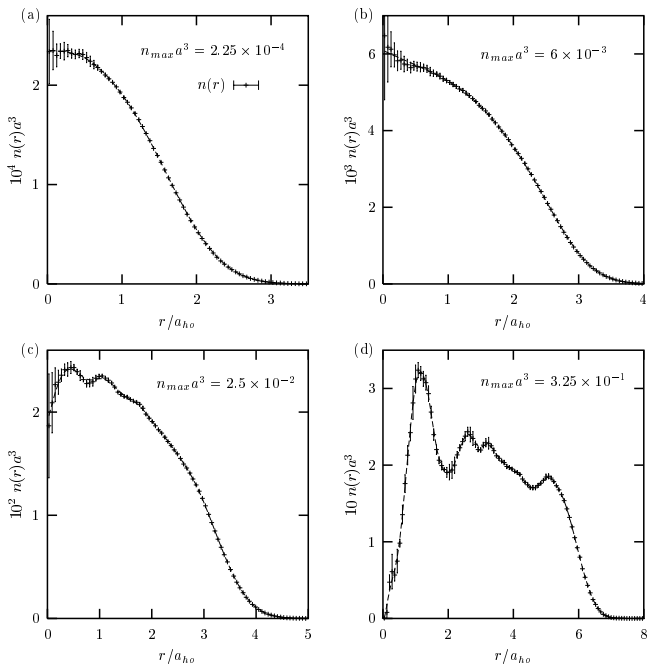} }
 \caption{DMC density profiles for hard sphere
trapped Bosons for four values of the maximum trap density $na^3$.
All plots are for $N=128$ and values of $a/a_{ho} =
0.03464,0.13856,0.27712,1.1084$ for frames (a),(b),(c), and (d)
respectively.} \label{DMC-density-comparison.eps}
\end{figure}

Figure \ref{DMC-density-comparison.eps} shows the DMC density
profiles for $128$ hard sphere trapped Bosons for four values of
the maximum trap density $na^3$. Frame (a) shows the radial
density profile for $a/a_{ho} =  8\times a_{Rb}/a_{ho} = 0.03464$.
The maximum density of the trapped Bosons in this system occurs at
the center of the trap with $na^3 \approx 2.25 \times 10^{-4}$.
This corresponds to a typical density observed in experiments in
metastable He$^*$ \cite{Robert2001a,DosSantos2001a}.  Frame (b)
shows results for $a/a_{ho} =  32\times a_{Rb}/a_{ho} = 0.13856$
and maximum density $na^3 \approx 6 \times 10^{-3}$, which is
comparable with densities found in $^{85}$Rb experiments.  In
frame (c), $a/a_{ho} = 64\times a_{Rb}/a_{ho} = 0.27712$.  Here,
local correlations in the density distribution near the center of
the trap are readily apparent. Finally, frame (d) shows the
density profile for $a/a_{ho} = 256\times a_{Rb}/a_{ho} = 1.1084$.
In this system, the hard spheres appear to have solidified in the
center of the trap. The $n(r)a^3$ for this system is only
qualitatively correct as mixed estimator bias caused by the
guiding function used is a factor here.

\subsubsection{Comparison of DMC $n(r)$ and Thomas-Fermi $n_{\rm
TF}(r)$}

\begin{figure}
\centerline{
    \includegraphics[width=.49\textwidth,bb=77 400 474 726,clip]{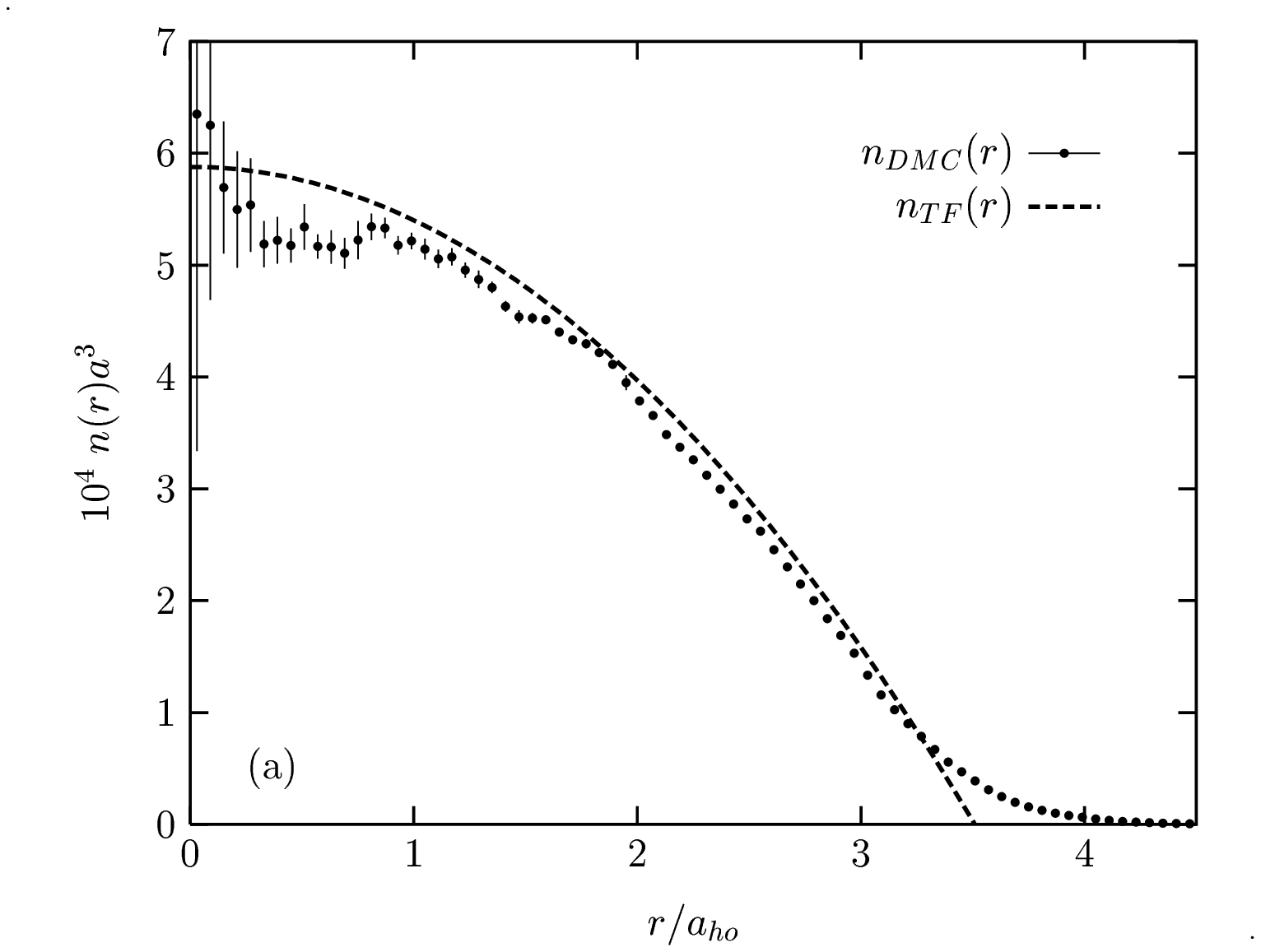}
} \centerline{
    \includegraphics[width=.49\textwidth,bb=77 400 474 726,clip]{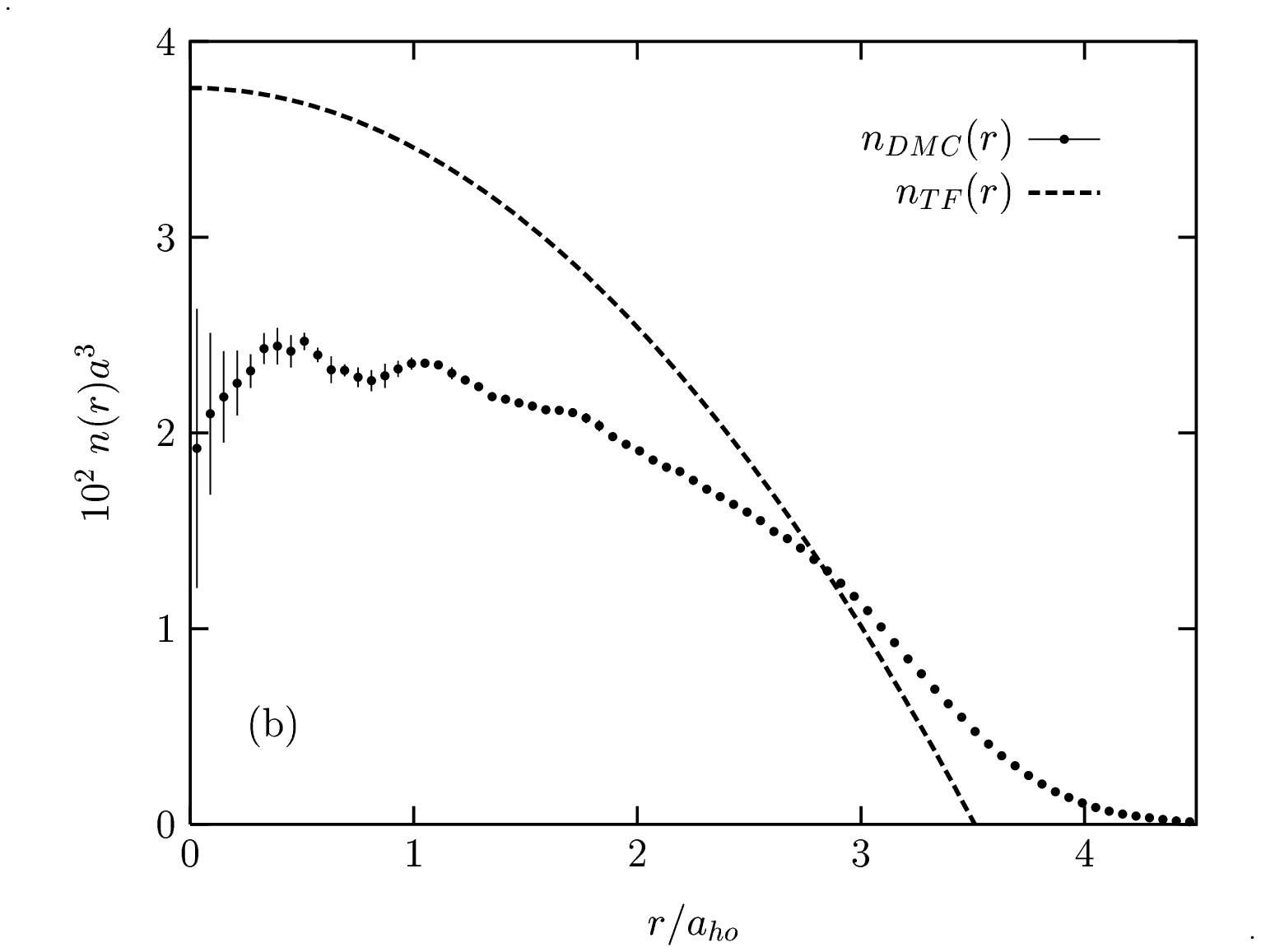}
}
    \caption{Comparison of total density distribution calculated
using diffusion Monte Carlo, $n_{\rm DMC}(r)$, for hard sphere
Bosons in a harmonic trap to the density predicted by the
Thomas-Fermi approximation (\ref{n(r)-TF}). Top frame (a) is for
$N=1024$ Bosons with ratio of scattering length to trap length of
$a/a_{ho} = 8 \times a_{Rb}~=~0.03464$. Density is expressed in
terms of $n(r)a^3 \times 10^4$. Frame (b) is for $N=128$ Bosons
with $a/a_{ho} = 64\times a_{Rb} = 0.27712$. Density is expressed
in terms of $n(r)a^3 \times 10^2$.
    } \label{DMC-vs-TF-rho.eps}
\end{figure}

In the so called ``Thomas-Fermi" (TF) approximate form of the
Gross-Pitaevskii equation, the interaction term $g \propto Na$ in
the GP equation is assumed to dominate the ``kinetic energy" or
gradient term resulting in an analytically solvable form of the GP
equation.  The TF approximation is expected to be valid when $Na$
is large, $Na >>1$, the interaction density is low, $na^3 <<1$,
and the ratio of the scattering length to the characteristic
harmonic trap length is small, $a/a_{ho} <<1$.

Figure \ref{DMC-vs-TF-rho.eps} shows a comparison of the total
density distribution calculated using DMC, $n_{\rm DMC}(r)$, to
the density predicted by the Thomas-Fermi approximation,
    \begin{equation}
    n_{\rm TF}(r) = ((15Na)^{2/5}-x^2)/8\pi Na. \label{n(r)-TF}
    \end{equation}
The top frame, (a), shows the density profile for $N=1024$ Bosons
with $a/a_{ho} = 8 \times a_{Rb}~=~0.03464$. Here, the TF and DMC
results agree quite well. However, the TF result slightly
overestimates the density near the center of the trap and fails to
reproduce the low density tail which occurs near the edge of the
trapped cloud. Frame (b) shows $n(r)$ for $N=128$ and $a/a_{ho} =
64\times a_{Rb} = 0.27712$. Note that the product, $Na/a_{ho}$,
(the only variable responsible for determining the shape of the TF
and GP density profiles) is the same in both frames. Clearly, in
the bottom frame, the TF approximation dramatically overestimates
the density at the center of the trap and underestimates the width
of the condensate.

\subsection{Condensate fraction}
\label{DMC_Condensate_Fraction_Section}
\begin{figure}
 \centerline{\includegraphics[width=.49\textwidth,bb= 153 310 552 659,clip]{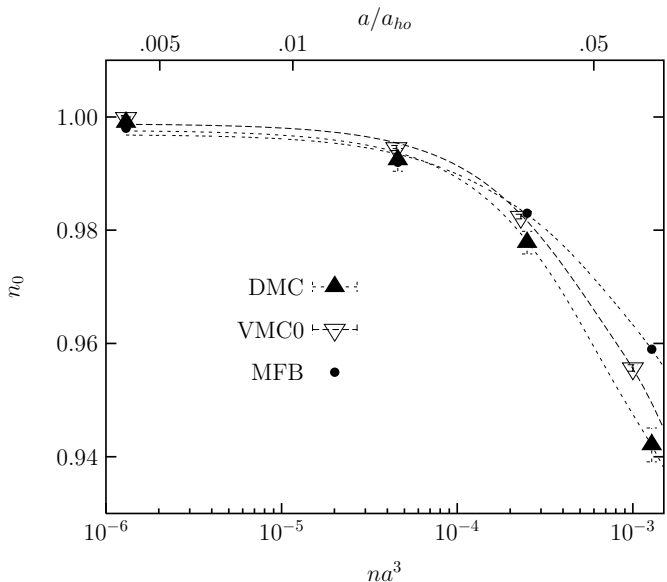}}
 \caption{ Condensate fraction, $n_0$, as a function of the
density, $na^3$, for $N=128$ trapped hard sphere Bosons. Here $n$
is the number density at the center of the trap and $a$ is the
scattering length. Circles are from the mean-field Bogoliubov
(MFB) expression for $n_0$ in a uniform dilute Bose gas integrated
over the TF density. The up and down-facing triangles are the DMC
and VMC results respectfully. Dashed lines are spline fits to
guide the eye.} \label{n0-dilute.eps}
\end{figure}

Figure \ref{n0-dilute.eps} shows the condensate fraction, $n_0$,
as a function of the density $na^3$ for $N=128$ trapped hard
sphere Bosons.  The $n$ is the maximum number density which is at
the center of the trap in the density range shown. The density was
varied by changing the value of $a/a_{ho}$. The corresponding
values of $a/a_{ho}$ are shown on the top axis. Circles are the
mean-field Bogoliubov (MFB) result for a uniform dilute Bose gas
integrated over the GP density in the Thomas-Fermi limit
\cite{Javanainen1996a} obtained by solving
\begin{equation}
n_0 = 1 - 0.3798(N_0a/a_{ho})^{{6}/{5}}/N. \label{MFB-n0}
\end{equation}

The up and down-facing triangles are the DMC and VMC0 results,
respectively, obtained from diagonalizing the OBDM. For  $na^3 <
10^{-4}$, all three values of $n_0$ agree to within $1\%$. At
higher densities, the MFB result consistently overestimates the
condensate fraction. MFB overestimates the condensate fraction
because it ignores local pair correlations which act to deplete
the condensate.  In contrast the DMC value of the condensate is
consistently lower than either the VMC0 or MFB estimates.  We
believe that the DMC result for $n_0$ is lower than either VMC and
MFB because it is able to treat local pair correlations more
accurately.  Pair correlations allow the total energy to decrease
at the expense of long range order. Since DMC is able to sample
the exact ground state, the mixed estimate for $n_0$ obtained from
DMC is more accurate than VMC or MFB.

\begin{figure}
\centerline{\includegraphics[width=.49\textwidth,bb= 163 309 553
672,clip ]{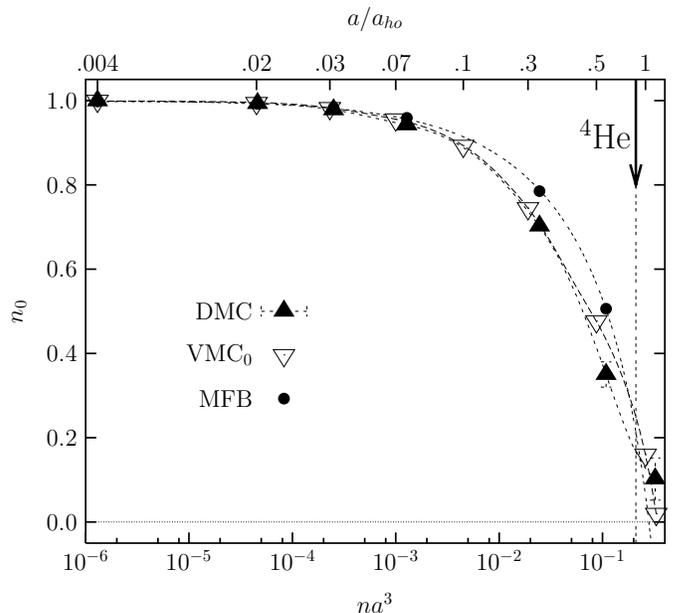}} \caption{Condensate fraction,
$n_0$, over a wide density range. The legend is the same as Fig.
\ref{n0-dilute.eps} } \label{n0-vs-na3.eps}
\end{figure}

Figure \ref{n0-vs-na3.eps} shows $n_0$ over a wider density range.
Here $n$ is again the maximum number density in the trap. At high
densities the maximum density in the trap is not always at the
center of the trap. As in the dilute regime presented in
Fig.~\ref{n0-dilute.eps}, MFB consistently overestimates the
condensate fraction for most densities.  At $na^3 \approx 0.28$,
however, the MFB estimate of $n_0$ goes to zero while both VMC and
DMC still show a condensate fraction of $n_0 \approx 10\%$.  The
MFB estimate goes to zero because the TF density profile used to
calculate the MFB value of $n_0$ does not have a broad low density
region near the surface of the trapped cloud of atoms as do the
DMC and VMC density distributions.  As will be shown in the next
section, the dilute region at the ``edge" of the trapped cloud can
support a condensate even when the condensate fraction in the
dense center of the trap goes to zero.

\begin{table}
\caption{Condensate fraction as obtained from
mean-field-Bogoliubov (MFB) , VMC, and DMC
 methods.}
\begin{ruledtabular}
\begin{tabular}{cllll}
{$na^3$} & $a/a_{ho}$ & {MFB} & {VMC} & {DMC}  \\
\hline
$1.3\times10^{-6}$ & 1 & 0.998 & 0.999(9) &  0.99(9) \\
$4.6\times10^{-5}$ & 4 & 0.992 & 0.992(4) & 0.99(2) \\
$2.5\times10^{-4}$ & 8 & 0.983 & 0.977(8) & 0.97(7) \\
$2.5\times10^{-3}$ & 16 & 0.959 & 0.942(1) & 0.94(2) \\
$2.4\times10^{-2}$ & 64 & 0.785 & 0.745(7) & 0.70(2)  \\
$1.1\times10^{-1}$ & 128 & 0.506 & 0.476(5) & 0.3(5) \\
$3.2\times10^{-1}$ & 256 & N/A\footnote{MFB predicts a negative condensate fraction for this system.}  & 0.160(0) & 0.1(0) \\
\end{tabular}
\end{ruledtabular}
\label{n0-table}
\end{table}

The density corresponding to liquid helium at SVP ($na^3 = 0.21$)
is indicated on the plot.  At this density, VMC gives a condensate
fraction of $n_0 \approx 25\%$ while DMC estimates a condensate
fraction of $n_0 \approx 18\%$.  In bulk liquid $^4$He, the
condensate fraction is $n_0 \approx 7.25\%$~\cite{Glyde2000a}.
This difference is explained by the fact that the dilute
region near the surface of the trapped cloud allows for a larger
fraction of particles to occupy the condensate orbital than in an
uniform system at $^4$He densities.

Table \ref{n0-table} summarizes the present DMC and VMC results
for the condensate fraction over a wide density range.

\subsection{Spatially dependent depletion of the condensate}

\begin{figure}
\centerline{
    \includegraphics[width=.49\textwidth,bb = 75 400 475 721,clip]{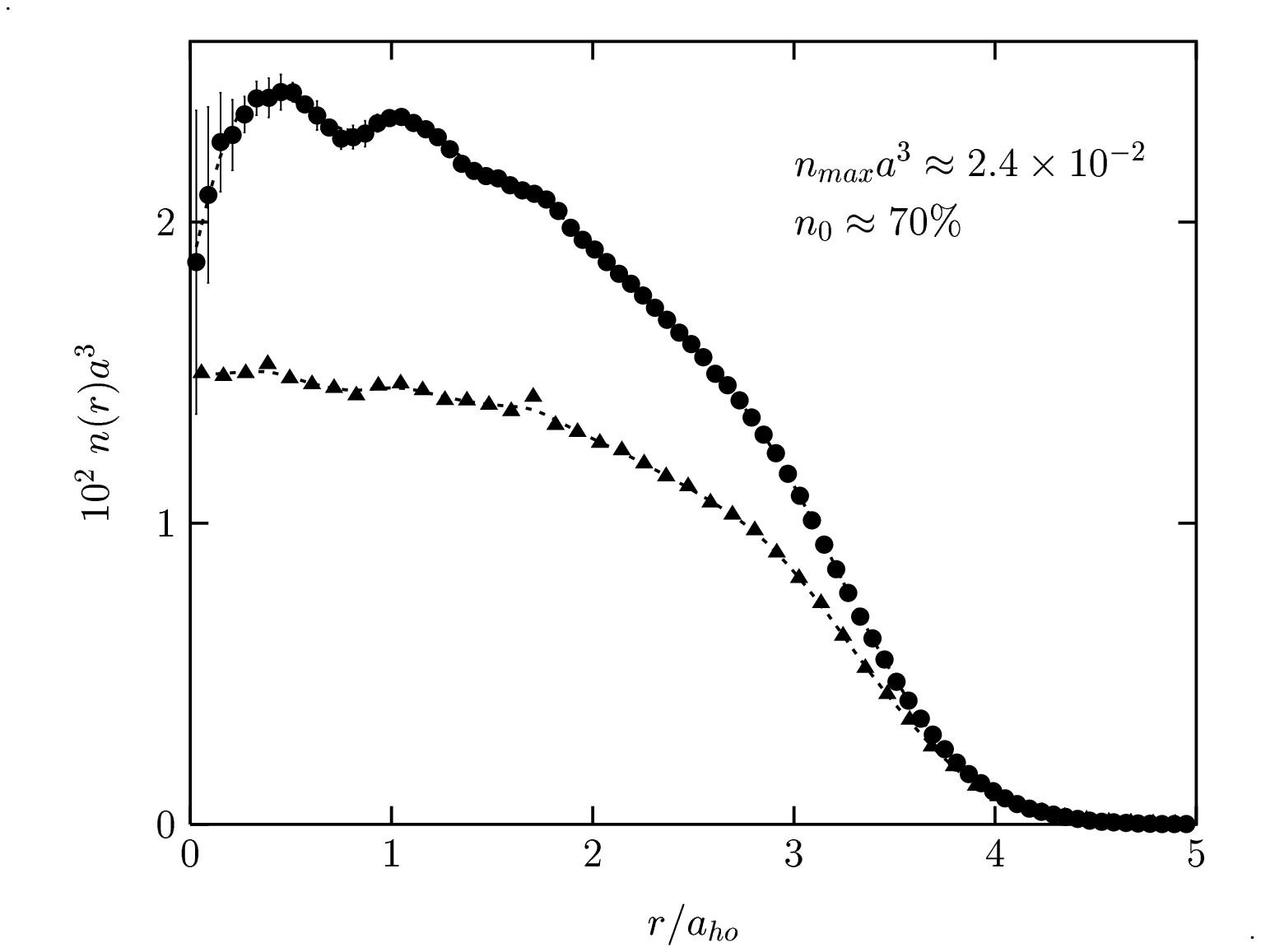}
}
\centerline{
    \includegraphics[width=.49\textwidth,bb = 75 400 475 721,clip]{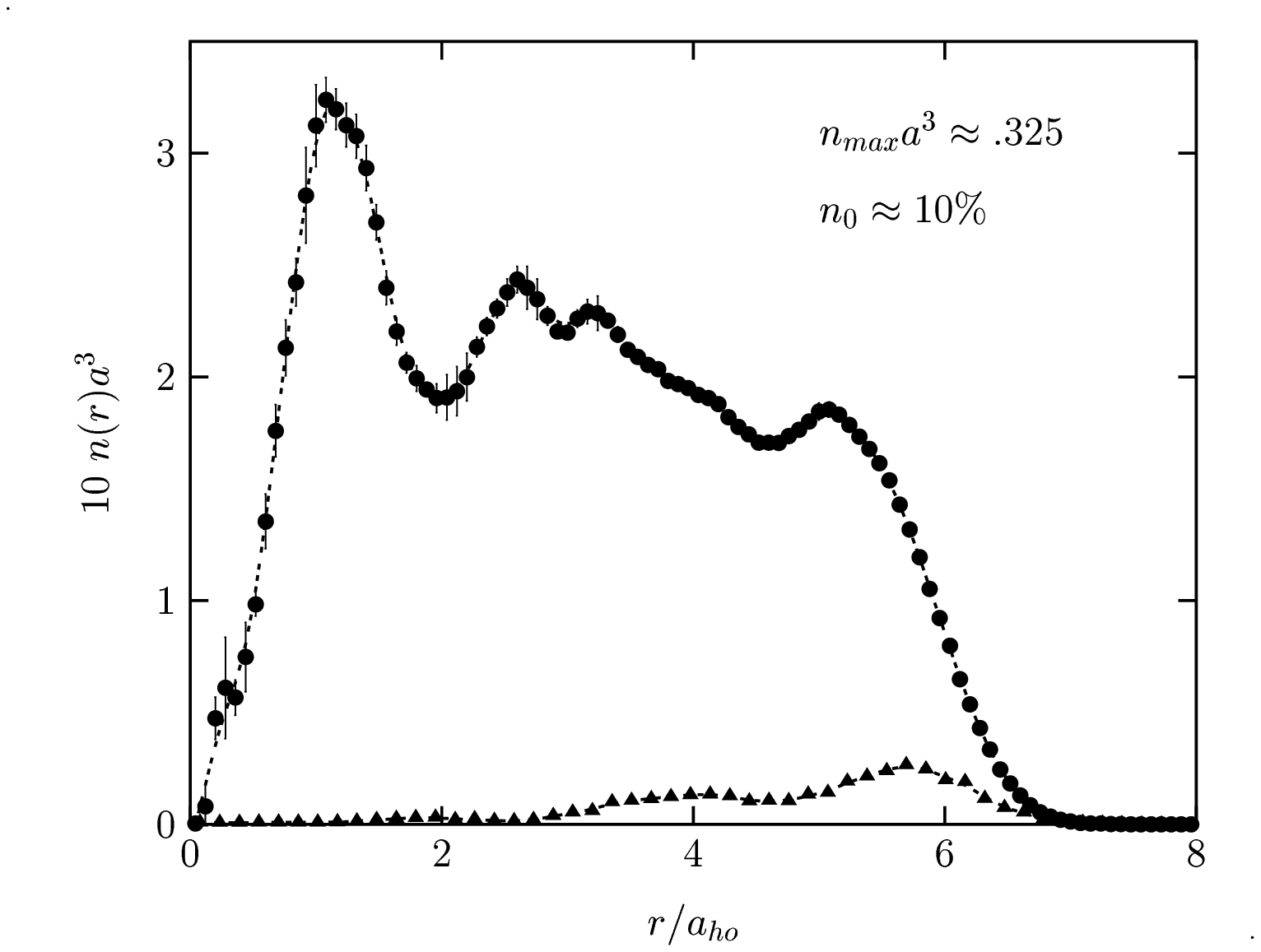}
}
    \caption{Comparison of total density distribution, $n(r)$, to condensate
    distribution, $n_0(r) = |\phi(r)|^2$,
    for $N=128$ hard sphere Bosons in a harmonic trap calculated using diffusion Monte Carlo.
    Circles are the total density while triangles represent the condensate. Dashed lines
    are spline fits to guide the eye.  In the top frame, the maximum
    density in the trap is $na^3 \approx 2.4\times 10^{-2}$ and the total condensate fraction
    is $n_0 \approx 70\%$.  In the bottom frame, $na^3 \approx 2.4\times 10^{-2}$
    and $n_0 \approx 10\%$.} \label{n0-vs-rho.eps}
\end{figure}

In Fig. \ref{n0-vs-rho.eps} we compare the total density
distribution, $n(r)$, to the condensate distribution, $n_0(r) =
n_0|\phi_0(r)|^2$, for $N=128$ hard sphere Bosons in a harmonic
trap calculated using diffusion Monte Carlo.  In the top frame,
$a/a_{ho} = 64 \times a_{Rb}/a_{ho}$ giving a maximum density in
the trap of $na^3 \approx 2.4\times 10^{-2}$ and total condensate
fraction of $n_0 \approx 70\%$.  In this system, the spatial
distribution of the condensate follows the shape of the total
density distribution except at small $r$. It is worth noting that
while the total density exhibits local correlations in the dense
region near the center of the trap, the condensate distribution is
relatively flat in this region. In the bottom frame of the figure,
$a/a_{ho} = 256\times a_{Rb}/a_{ho} = 1.1084$ resulting in a
maximum density of $na^3 \approx 0.325$ and a condensate fraction
of $n_0 \approx 10\%$. This is the same system shown in frame~(d)
of Fig. \ref{DMC-density-comparison.eps}.  As discussed above, the
DMC results at this density are biased by the VMC guiding function
used.  Nevertheless, we believe the results to be qualitatively
correct.   Here, strong pair correlations have completely depleted
the condensate in the center of the trap but the relatively dilute
region near the edge of the trap is still able to support a
condensate.  We find that for trapped hard sphere Bosons, the
local condensate fraction, $n_0(r)/n(r)$, rises in the dilute
region near the surface and remains close to one all the way to
the surface of the cloud. This may be contrasted with predictions
for $n_0(r)/n(r)$ for self-bound superfluid $^4$He at a free
surface in which surface correlations significantly deplete the
condensate at the liquid-vacuum interface
\cite{Galli2000a,Draeger2002a}.

\section{Discussion}
\label{discussion section}
 The main objectives of this work are to
explore the role of interactions in determining the zero
temperature properties of the trapped Bose gas over a wide range
of densities and to determine the limits of the mean-field
description of the condensate properties.   To this end, we have
employed quantum Monte-Carlo (QMC) methods and the one-body
density matrix (OBDM) formulation of BEC.  We find the OBDM
description of a many-body Bose system combined with QMC
techniques, provides a coherent method for the study of the ground
state properties and Bose-Einstein condensation in traps from the
dilute to the very dense regime.  By comparing our QMC results
with mean-field theory we determine key limits of the mean field
description.

\subsection{The ground state energy}
We find that in the dilute limit, $na^3 \lesssim 10^{-4}$, where
the condensate depletion is small, $n_0\gtrsim 99\%$, the GP
description of the condensate provides a good description of the
full many-body ground state. Once the density has reached,
$na^3\approx 10^{-3}$, approximately $6\%$ of the atoms lie
outside of the condensate and the condensate energy obtained from
GP theory lies $3\%$ below the QMC energy.  For $na^3 \gtrsim
10^{-3}$, the GP energy does not describe the energy of the Bosons
in the trap accurately. The present QMC corrections to the GP
energy, $\delta E = E_{DMC}-E_{GP}$, are proportional to N$^{3/5}$
when N is allowed to vary with fixed $a$ and are proportional to
$(a/a_{ho})^{8/5}$ when $a$ is allowed to vary with fixed N.  This
dependence on N and $a$ holds for all densities studied ($10^{-6}
< na^3 < 0.5$) and is consistent with the expected corrections to
the GP energy arising from the depletion of the
condensate~(\ref{dE-LDA}). Thus, the GP description of the
condensate energy appears to be valid even in the highly
interacting regime. However, the dependence of $\delta E$ on the
product $N^{3/5}(a/a_{ho})^{8/5}$, as predicted by MGP
(\ref{dE-LDA}), holds up to densities $na^3\approx 5 \times
10^{-4}$ only. As interaction is increased the effects of the
non-condensate play an increasingly significant roll in
determining the properties of the total ground state and a more
complicated functional dependence of $\delta E(N,a)$ than the
simple product $N^{3/5}(a/a_{ho})^{8/5}$ is required at higher
densities.

 \subsection{Deviations from the mean-field description}

\begin{figure}
\includegraphics[width=.49\textwidth,bb= 87 400 473 734,clip]{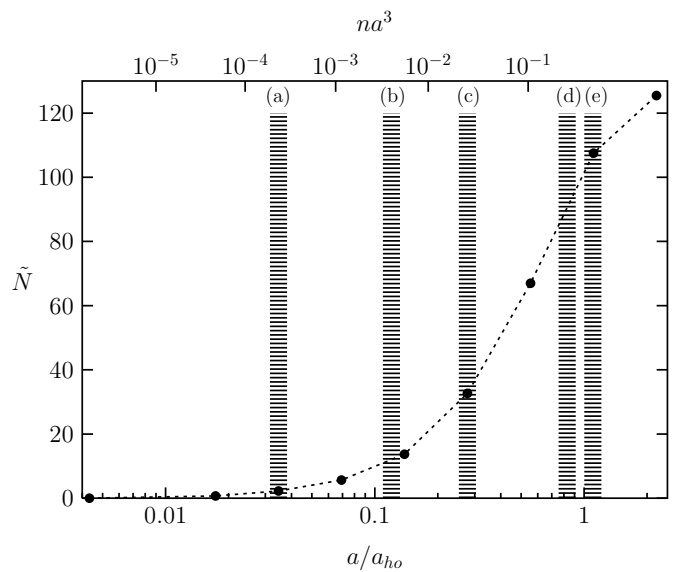}
\caption{QMC determination of the number of Bosons outside the
condensate, $\tilde{N}$, for $N=128$ hard sphere Bosons in a
spherically symmetric harmonic trap vs. $a/a_{ho}$. The
corresponding densities, $na^3$, are shown across the top axis.
Striped bands indicate regions in $a/a_{ho}$ where the QMC results
for BEC properties diverge from mean-field / Bogoliubov
predictions. (a) QMC and Bogoliubov results for $n_0$ begin to
diverge. (b) QMC and mean-field results for the size of the
condensate diverge. (c) Local correlations in the density profile
of the many-body ground state begin to appear. (d) Condensate
begins to shift from center of trap. (e) Condensate exists only in
dilute region near surface of trapped cloud. }
\label{critical_n0.eps}
\end{figure}

Figure \ref{critical_n0.eps} contains striped bands indicating
regions in $a/a_{ho}$ where QMC results diverge from mean-field /
Bogoliubov predictions.  Since the degree of depletion of the
condensate arising from inter-Boson interaction plays a
significant role in determining beyond-mean-field effects, QMC
values for the number of atoms outside the condensate,
$\tilde{N}$, for a system with a total of $N=128$ Bosons are shown
along with the regions. The first sign of divergence (a) occurs at
a density of $na^3 \approx 3\times 10^{-4}$ and a value of
$a/a_{ho} \approx 8 \times a_{Rb}/a_{ho} \approx 0.035$. At this
density, QMC and MFB (\ref{MFB-n0}) results for the condensate
fraction, $n_0 = N_0/N$, begin to diverge (see
Fig.~\ref{n0-dilute.eps}.). Below this value of $a/a_{ho}$, QMC
and MFB values of $n_{0}$ agree to within $1\%$. At $na^3\approx
10^{-3}$, MFB underestimates the depletion of the condensate by
$30\%$. At higher densities, $na^3 \approx 10^{-1}$, MFB predicts
a condensate fraction $40\%$ higher than QMC.

The second point of interest in Fig. \ref{critical_n0.eps} marked
(b) occurs in the region of $na^3 \approx 2.5\times 10^{-3}$ and
$a/a_{ho} \approx  0.12$. Near this value of $a/a_{ho}$, QMC
results for the size of the many-body ground state and mean-field
results for the size of the condensate (\ref{TF-radius}) begin to
differ.  For values of $a/a_{ho}~\lesssim~0.12$, the width of the
many-body ground state is proportional to $(Na/a_{ho})^{1/5}$ as
predicted by mean-field theory. At higher values,$(a/a_{ho} >
0.12)$, we find that the size of the condensate is better
described by a scaling of $(a/a_{ho})^{2/3}$. The scaling is shown
in Fig. \ref{width-vs-Na.eps} and discussed in Section
\ref{DMC-vs-TF-Width}.   Thus, for systems with
$na^3~\gtrsim~10^{-3}$, GP theory in the TF limit under-estimates
the growth of the size of the ground state with $a/a_{ho}$
significantly. In the extreme range of very large scattering
length or very tight trapping potential where $a/a_{ho} = 1$, GP
predicts a condensate distribution $20\%$ smaller than the width
of the ground state obtained from QMC.

The band (c) in Fig. \ref{critical_n0.eps} indicates the region in
which local correlations in the density profile of the many-body
ground state begin to appear.  These local correlations signal a
clear departure from mean-field properties.  This effect occurs
for systems with trap densities of $na^3~\gtrsim~2.5\times
10^{-2}$ and $a/a_{ho} \approx 64\times a_{Rb}/a_{ho} \approx
0.28$. At this level of interaction the condensate fraction as
obtained from DMC  is $n_0 \approx 70\%$.  We find that at this
density the condensate density is smoothly varying throughout the
trap with little or no local density fluctuations (See top frame
of Fig. \ref{n0-vs-rho.eps}). Evidence that the condensate
distribution does not explicitly follow the total density
distribution is another demonstration that a local density
approximation description of the condensate breaks down at this
density.

The band marked (d) in Fig. \ref{critical_n0.eps} approximates the
region in which the condensate begins to shift from the center of
the trap to the surface.  Here, $na^3 \approx 0.2$ and $a/a_{ho}
\approx 0.8$.  The condensate fraction is $n_0 \approx 20\%$.  At
this level of interaction and beyond, mean-field approximations
and the Bogoliubov approximation both fail to appropriately
describe the properties of a trapped BEC. We speculate that at
this density, the increased depletion in the center of the trap
could effectively pin vortex states.

The final point of interest in Fig.~\ref{critical_n0.eps} occurs
in the region marked by the band (e).  For systems with
$na^3~\gtrsim~0.3$ and $a/a_{ho}~\gtrsim~256\times a_{Rb}/a_{ho}
\approx 1.1$, the condensate exists only in dilute region near
surface of trapped cloud. Strong pair correlations have completely
depleted the condensate in the center of the trap but the
relatively dilute region near the edge of the trap is still able
to support a condensate.  Figure \ref{n0-vs-rho.eps} presents DMC
results which demonstrate this phenomena.

\acknowledgments {Stimulating discussions with J. Boronat and S.
A. Chin as well as comments on the original manuscript from D.
Blume are gratefully acknowledged. Partial support from National
Science Foundation DMR-0115663 is gratefully acknowledged. }

\bibliographystyle{apsrev}
\bibliography{bib_files/JILA-BEC,bib_files/mybib}

\end{document}